\documentclass{iopart}
\usepackage{amsfonts,amssymb,amstext}
\usepackage{cite}
\usepackage{graphicx}
\usepackage{subfig}

\begin{document}

\sloppy

\title[Electromagnetic signatures of gravitational radiation]{Electromagnetic signatures of far-field gravitational radiation in the 1+3 approach}
\author{Alvin J K Chua, Priscilla Ca\~{n}izares and Jonathan R Gair}
\address{Institute of Astronomy, University of Cambridge, Madingley Road, Cambridge CB3 0HA, United Kingdom}
\eads{\mailto{ajkc3@ast.cam.ac.uk}, \mailto{pcm@ast.cam.ac.uk}, \mailto{jgair@ast.cam.ac.uk}}

\begin{abstract}
Gravitational waves from astrophysical sources can interact with background electromagnetic fields, giving rise to distinctive and potentially detectable electromagnetic signatures. In this paper, we study such interactions for far-field gravitational radiation using the 1+3 approach to relativity. Linearised equations for the electromagnetic field on perturbed Minkowski space are derived and solved analytically. The inverse Gertsenshte\u\i n conversion of gravitational waves in a static electromagnetic field is rederived, and the resultant electromagnetic radiation is shown to be significant for highly magnetised pulsars in compact binary systems. We also obtain a variety of nonlinear interference effects for interacting gravitational and electromagnetic waves, although wave--wave resonances previously described in the literature are absent when the electric--magnetic self-interaction is taken into account. The fluctuation and amplification of electromagnetic energy flux as the gravitational wave strength increases towards the gravitational--electromagnetic frequency ratio is a possible signature of gravitational radiation from extended astrophysical sources.
\end{abstract}

\submitto{\CQG}

\maketitle

\section{Introduction}
Searches with pulsar timing arrays \cite{HEA2010}, ground-based detectors such as Advanced LIGO \cite{H2010}, and the proposed space-based mission eLISA \cite{AEA2012} are expected to begin yielding detections of gravitational waves (GWs) in the near future. The most promising GW sources anticipated for current and future detectors are highly energetic astrophysical events; these include supernovae, compact stellar-mass binaries and massive black hole mergers, many of which will be accompanied by distinctive and detectable electromagnetic signals. Such electromagnetic counterparts can aid ongoing GW detection efforts through improved event rate prediction, enhanced source parameter estimation, the provision of search triggers, and the identification and/or confirmation of individual detections. Once our ability to detect gravitational radiation is on a firm footing, the synergy of complementary information from gravitational and electromagnetic observations should establish GWs as an important component of multi-messenger astronomy \cite{P2009,MO2010,MB2012,PNR2013,GN2014,SEA2014}.

As observations of GW sources and their electromagnetic counterparts improve in precision, so too must models of such dual sources, to account for any correlations between the two types of signal. A nascent line of research towards this end is the direct coupling between gravitational and electromagnetic fields in the strong-field regime. Recent work in this area has focused on the electromagnetic signatures of gravitational perturbations on various curved spacetimes; the perturbed Einstein--Maxwell equations have been solved for Schwarzschild \cite{SKLS2013}, slowly rotating Kerr--Newman \cite{PBG2013} and equal-mass binary Kerr \cite{PEA2009}, with the numerical involvement increasing as per the complexity of the spacetime.

The problem of Einstein--Maxwell coupling for gravitational radiation in flat space is older and more analytically tractable than that in curved space, leading to a better characterisation of the (albeit weaker) interactions between far-field GWs and electromagnetic fields. One such effect is the resonant conversion of a GW into an electromagnetic wave (EMW)---and vice versa---in the presence of a static electromagnetic field \cite{G1962,Z1974,BH2010}. The direct signatures of GWs on EMWs have also been studied; these include frequency splitting \cite{C1968}, intensity fluctuations \cite{C1968,Z1966}, deflection of rays \cite{Z1966,F1993,KKP2006} and gravitationally induced rotation of the EMW polarisation \cite{KKP2006,C1983,M1998,PM2002,HG2007,F2008}. Indirect GW detection schemes using microlensing \cite{RVM2003} and phase modulation \cite{LLMM2005} effects on light have been proposed as well.

Among various frameworks suited to the study of interacting GWs and electromagnetic fields is the 1+3 covariant approach to general relativity, in which spacetime is locally split into time and space via the introduction of a fundamental timelike congruence \cite{E1971,EV1999,TCM2008}. This approach is most commonly employed in the cosmological setting, and in particular has been used to describe electromagnetic signatures of the tensor perturbations associated with cosmological GWs \cite{DBE1997,MDB2000,TDM2003,T2010,T2011}. It may also be applied to gravitational--electromagnetic interactions in a general spacetime \cite{T2005}, although any inhomogeneity in the spacetime typically renders the governing equations intractable due to tensor--vector and tensor--tensor coupling \cite{CB2003}.

Such difficulties with the 1+3 formalism may be partially overcome by extending the spacetime splitting to a 1+1+2 decomposition in the case of locally rotationally symmetric $G_2$ spacetimes, which have a preferred spatial direction \cite{CB2003}; this method has been used to semi-analytically model the electromagnetic signature of a Schwarzschild ringdown \cite{CMBD2004}. The 1+3 approach may also be supplemented by an orthonormal tetrad formalism \cite{EV1999}, which has been applied to the interaction of far-field GWs and electromagnetic fields in the presence of a magnetised plasma \cite{MBD2000,SB2003}. Finally, recent work on Minkowski-space GWs and EMWs within the 1+3 framework has uncovered resonant interactions between the two under specific conditions \cite{T2011,KT2013}.

As any resonant amplification of electromagnetic fields by gravitational radiation might be important in the context of GW detection, we take a more detailed look at flat-space interactions between GWs and electromagnetic fields within the 1+3 approach. In Sec. \ref{sec:interactions}, we derive linearised evolution and constraint equations for the electromagnetic field on GW-perturbed flat space, and approximate these on exact Minkowski space. This framework is applied to simple models of static and radiative electromagnetic fields in Sec. \ref{sec:signatures}, where we consider the resultant effects in astrophysical settings and discuss their implications for dual observations of GW sources.

We rederive in Sec. \ref{subsec:static} the resonant induction of an EMW by a GW in a static electromagnetic field, and estimate that for highly magnetised pulsars in compact binary systems, the energy radiated through this process might be non-negligible with respect to the magnetic dipole radiation. In Sec. \ref{subsec:radiation}, we find no resonant interaction between plane GWs and EMWs after considering electric--magnetic self-interaction contributions that have been omitted in previous work \cite{T2011,KT2013}. However, nonlinear interference effects are shown to be significant in a regime where the GW strength approaches the GW--EMW frequency ratio from below; the resultant fluctuation and amplification of electromagnetic energy flux is a potentially stronger signature of gravitational radiation than other geometrical-optics effects in the literature.

We use geometrised units $c=8\pi G=\mu_0=1$ in this paper. Latin (spacetime) indices run from 0 to 3, while Greek (space) indices run from 1 to 3; the metric signature is $(-,+,+,+)$ and the Riemann tensor sign convention is $R_{ab}=R^c_{\enspace acb}$.

\section{Far-field gravitational--electromagnetic interactions}\label{sec:interactions}
In the 1+3 covariant approach to general relativity, we introduce a timelike vector field $u^a$ tangential to a congruence of world lines on a general spacetime. This fundamental four-velocity field is normalised such that $u_au^a=-1$, and in the absence of vorticity foliates the spacetime into spacelike hypersurfaces orthogonal to $u^a$. Every quantity on the spacetime may then be decomposed into its timelike and spacelike parts; in addition, the covariant time ($\dot{\mathcal{T}}$) and space ($\mathcal{T}_{:a}$) derivatives of a tensor field $\mathcal{T}$ are defined as the respective projections of its covariant derivative $\mathcal{T}_{;a}$ tangential and orthogonal to $u^a$ \cite{TCM2008}. If $\mathcal{T}$ is a spatially projected tensor, we may write
\begin{equation}
\dot{\mathcal{T}}:=\mathcal{T}_{;a}u^a,\quad\mathcal{T}_{:a}:=\mathcal{T}_{;a}+\dot{\mathcal{T}}u_a.
\end{equation}

We consider the interactions between gravitational radiation and electromagnetic fields in perturbed Minkowski space with the metric $\tilde{\eta}_{ab}$. In the transverse--traceless gauge, $\tilde{\eta}_{00}=-1$ and we may choose $u^a=\delta^a_0$, where $\delta^a_b$ is the Kronecker delta. Gravitational radiation is covariantly described by the transverse electric ($E_{ab}$) and magnetic ($H_{ab}$) parts of the Weyl tensor, and more simply in flat space by the shear tensor $\sigma_{ab}$ (the traceless part of $u_{(a:b)}$), which satisfies the transversality condition $\tilde{\mathrm{div}}\sigma_{ab}:=\sigma_{ab}^{\quad:b}=0$ to linear perturbative order \cite{TCM2008}. The shear is related to the usual transverse--traceless metric perturbation by
\begin{equation}\label{eq:metric}
\sigma_{ab}=\frac{1}{2}\dot{h}^\mathrm{TT}_{ab}.
\end{equation}
Other kinematical and geometrical quantities are greatly simplified by the flatness and symmetry of the spacetime. The acceleration and vorticity are identically zero, while at linear order the expansion $\vartheta=u^a_{\enspace:a}$, spatially projected three-Ricci tensor $\mathcal{R}_{ab}$ and Weyl tensor components reduce to \cite{TCM2008}
\begin{equation}\label{eq:linear}
\vartheta=0,\quad\mathcal{R}_{ab}=E_{ab}=-\dot{\sigma}_{ab},\quad H_{ab}=\tilde{\mathrm{curl}}\sigma_{ab},
\end{equation}
where $\tilde{\mathrm{curl}}\sigma_{ab}:=\epsilon_{cd(a}\sigma_{b)}^{\enspace d:c}$ and $\epsilon_{\mu\nu\rho}$ is the three-dimensional Levi-Civita symbol.$^1$\footnote[0]{$^1$Here we use $\epsilon_{0123}=(-\mathrm{det}\tilde{\eta}_{ab})^{1/2}$ for the spacetime volume form and $\epsilon_{abc}=u^d\epsilon_{dabc}$ for its spatial projection \cite{EV1999}, such that $\epsilon_{123}=1$ to linear perturbative order.}

Maxwell's equations may likewise be decomposed in the 1+3 formalism, allowing the derivation of wave-like equations for the electromagnetic field. The exact first-order equations for the spatially projected electric ($E_a$) and magnetic ($B_a$) fields on our spacetime are \cite{TCM2008}
\begin{equation}\label{eq:div}
\tilde{\mathrm{div}}E_a=0,\quad\tilde{\mathrm{div}}B_a=0,
\end{equation}
\begin{equation}\label{eq:curl}
\tilde{\mathrm{curl}}E_a=-\dot{B}_a+\sigma_{ab}B^b-\frac{2}{3}\vartheta B_a,\quad\tilde{\mathrm{curl}}B_a=\dot{E}_a-\sigma_{ab}E^b+\frac{2}{3}\vartheta E_a,
\end{equation}
where $\tilde{\mathrm{div}}V_a:=V_a^{\enspace:a}$ and $\tilde{\mathrm{curl}}V_a:=\epsilon_{abc}V^{c:b}$ for any spatially projected vector $V_a$. We obtain second-order evolution equations for $E_a$ and $B_a$ by spatially projecting the covariant time derivatives of Eqs \eref{eq:curl}, making use of the Ricci and Bianchi identities as well as the Raychaudhuri equation for $\dot{\vartheta}$. These wave-like equations for $E_a$ and $B_a$ are sourced by the kinematical quantities in \eref{eq:curl}, with additional Ricci and Weyl curvature terms arising from the non-commutativity of derivatives. Similar equations have been derived for a fully general spacetime, where the only assumption is a single perfect-fluid matter field with a barotropic equation of state \cite{T2005}.

For most astrophysical GW sources we expect to observe, the electromagnetic luminosity ($\sim10^{37}\,\mathrm{W}$ for a typical galaxy) is dwarfed by the gravitational luminosity (some significant fraction of $c^5/G\sim10^{52}\,\mathrm{W}$) \cite{SS2009}, and so the energy carried by gravitational radiation is generally much greater than that stored in the electromagnetic field. This translates to $E^2\sim B^2\ll\sigma^2\ll1$ in our units, where $E^2:=E^aE^*_a$, $B^2:=B^aB^*_a$ and $\sigma^2:=\sigma^{ab}\sigma^*_{ab}/2$. The evolution equations for $E_a$ and $B_a$ contain source terms of three sizes: $\sim\sigma E$, $\sim\sigma^2E$ and $\mathcal{O}(E^3)$, with the last arising from the back-reaction of the electromagnetic field on the background spacetime via the Einstein field equations. Considering only the leading (in $\sigma$) terms at linear order in $E$ and using the linearised relations \eref{eq:linear}, we write
\begin{equation}\label{eq:boxE}
\tilde{\Box}E_a=\sigma_{ab}\dot{E}^b+2\dot{\sigma}_{ab}E^b+\epsilon_{abc}\sigma^{cd}B_d^{\enspace:b}+\epsilon_{abc}\sigma^{cd:b}B_d+(\tilde{\mathrm{curl}}\sigma_{ab})B^b,
\end{equation}
\begin{equation}\label{eq:boxB}
\tilde{\Box}B_a=\sigma_{ab}\dot{B}^b+2\dot{\sigma}_{ab}B^b-\epsilon_{abc}\sigma^{cd}E_d^{\enspace:b}-\epsilon_{abc}\sigma^{cd:b}E_d-(\tilde{\mathrm{curl}}\sigma_{ab})E^b,
\end{equation}
where $\tilde{\Box}\mathcal{T}:=\ddot{\mathcal{T}}-\mathcal{T}_{:a}^{\enspace:a}$ for any spatially projected tensor $\mathcal{T}$.

Eqs \eref{eq:boxE} and \eref{eq:boxB}, along with the divergence constraints \eref{eq:div}, govern the evolution of electromagnetic fields in the presence of far-field gravitational radiation. They are coupled to the usual first-order propagation and constraint equations for $\sigma_{ab}$ \cite{TCM2008}, which may be cast as a constrained wave-like equation in similar fashion to the derivation of \eref{eq:boxE} and \eref{eq:boxB}. The shear equations contain terms that are $\sim\sigma^2$, $\sim\sigma^3$ and $\mathcal{O}(E^2\sigma)$; at linear order in $\sigma$, however, we have
\begin{equation}\label{eq:boxsigma}
\tilde{\mathrm{div}}\sigma_{ab}=0,\quad\tilde{\Box}\sigma_{ab}=0.
\end{equation}
Hence it is reasonable to treat $\sigma_{ab}$ as a fixed background of gravitational radiation that drives oscillations in the electromagnetic field via \eref{eq:boxE} and \eref{eq:boxB}.

For far-field calculations, it is convenient to replace the perturbed Minkowski metric $\tilde{\eta}_{ab}$ with an exact one $\bar{\eta}_{ab}$, which simplifies index manipulation and any harmonic expansion of tensor fields. This approximation is trivially valid for Eqs \eref{eq:boxsigma}, where replacing the covariant divergence and d'Alembert operators with their Minkowski counterparts only introduces terms that are quadratic- or higher-order in $\sigma$, but not so for \eref{eq:boxE} and \eref{eq:boxB}. Using a perturbative approach, we consider the gravitationally coupled electromagnetic field as the sum of a free field and an induced first-order perturbation, i.e.
\begin{equation}\label{eq:perturb}
E_a=E^{(0)}_a+E^{(1)}_a,\quad B_a=B^{(0)}_a+B^{(1)}_a,
\end{equation}
where $\{E^{(0)}_a,B^{(0)}_a\}$ is a vacuum Maxwell solution, $E^{(1)}\ll E^{(0)}$ and $B^{(1)}\ll B^{(0)}$. Denoting the divergence and d'Alembert operators on Minkowski space as $\bar{\mathrm{div}}$ and $\bar{\Box}$ respectively, we have
\begin{equation}\label{eq:Maxwell}
\bar{\mathrm{div}}E^{(0)}_a=0,\quad\bar{\mathrm{div}}B^{(0)}_a=0,\quad\bar{\Box}E^{(0)}_a=0,\quad\bar{\Box}B^{(0)}_a=0.
\end{equation}

Substituting \eref{eq:perturb} and \eref{eq:Maxwell} into the equations for $\{E_a,B_a\}$ yields wave-like equations for the induced field that are essentially \eref{eq:boxE} and \eref{eq:boxB} with linear corrections. These corrections are due to the difference operators $(\tilde{\mathrm{div}}-\bar{\mathrm{div}})$ and $(\tilde{\Box}-\bar{\Box})$ giving rise to terms that are $\sim\sigma E$ and non-negligible with respect to \eref{eq:boxE} and \eref{eq:boxB}.
The induced field equations read
\begin{equation}\label{eq:divEcorr}
\bar{\mathrm{div}}E^{(1)}_a=\{\textrm{linear corrections}\}[E^{(0)}_a],
\end{equation}
\begin{equation}\label{eq:divBcorr}
\bar{\mathrm{div}}B^{(1)}_a=\{\textrm{linear corrections}\}[B^{(0)}_a],
\end{equation}
\begin{equation}\label{eq:boxEcorr}
\bar{\Box}E^{(1)}_a=F[E^{(0)}_a]+G[B^{(0)}_a]+\{\textrm{linear corrections}\}[E^{(0)}_a],
\end{equation}
\begin{equation}\label{eq:boxBcorr}
\bar{\Box}B^{(1)}_a=F[B^{(0)}_a]-G[E^{(0)}_a]+\{\textrm{linear corrections}\}[B^{(0)}_a].
\end{equation}
Here $F$ and $G$ are linear maps defined by \eref{eq:boxE} and \eref{eq:boxB} as
\begin{equation}\label{eq:F}
F[V_a]:=\sigma_{ab}\dot{V}^b+2\dot{\sigma}_{ab}V^b,
\end{equation}
\begin{equation}\label{eq:G}
G[V_a]:=\epsilon_{abc}\sigma^{cd}V_d^{\enspace:b}+\epsilon_{abc}\sigma^{cd:b}V_d+(\bar{\mathrm{curl}}\sigma_{ab})V^b,
\end{equation}
with the covariant time and space derivatives equal to their partial counterparts at linear perturbative order.

The divergences of the induced electromagnetic field contain terms that are generally nonzero, even in the absence of sources. Eq. \eref{eq:divEcorr} in particular has been interpreted as an effective four-current generator for the induced field \cite{C1968}, although there is no similar analogy for its magnetic counterpart \eref{eq:divBcorr}. A more suitable comparison might be to think of the corrections in \eref{eq:divEcorr} and \eref{eq:divBcorr} as ``polarisation'' and ``magnetisation'' effects generated by the spacetime perturbations, with $E_a$ and $B^{(0)}_a$ playing the respective roles of the electric displacement and auxiliary magnetic fields \cite{J1998}.

In this paper, we consider a background GW that is plane, monochromatic and linearly polarised with constant amplitude. The geometrical-optics approximation is valid whenever the gravitational wavelength is much shorter than the background radius of curvature, i.e. across the distant wave zone of a typical astrophysical source and well into its local wave zone \cite{T1980}. More realistic (multimodal) inspiral-type waveforms for the time-varying part of the GW may be built up from superpositions of our simplified model, with the resultant imprint on the electromagnetic field bearing the characteristics of the source waveform.

The GW is governed by the shear equations \eref{eq:boxsigma} with $\tilde{\mathrm{div}}=\bar{\mathrm{div}}$ and $\tilde{\Box}=\bar{\Box}$. In coordinates $x^a=(t,x,y,z)$ such that it propagates in the $z$-direction with zero initial phase, the wave is described by the real part of
\begin{equation}\label{eq:sineGW}
\sigma_{ab}=\sigma\exp{(-ik(t-z))}p_{ab},
\end{equation}
with spatial wave vector $k_\mu=k\delta^3_\mu$. The unit polarisation tensor $p_{ab}$ has nonzero components $p_{11}=-p_{22}=\cos{2\alpha}$ and $p_{12}=p_{21}=\sin{2\alpha}$ for some wave polarisation angle $\alpha$. Any linear corrections in Eqs \eref{eq:divEcorr}--\eref{eq:boxBcorr} are then obtained in the usual way with the metric perturbation, which is given by the real part of
\begin{equation}
h^\mathrm{TT}_{ab}=\frac{2i}{k}\sigma_{ab},
\end{equation}
in accordance with \eref{eq:metric}.

\section{Electromagnetic signatures of gravitational radiation}\label{sec:signatures}
We now consider two simple models for the free field $\{E^{(0)}_a,B^{(0)}_a\}$ in Eqs \eref{eq:divEcorr}--\eref{eq:boxBcorr}, and discuss their astrophysical implications. Sec. \ref{subsec:static} deals with the effects of gravitational radiation on a static electromagnetic field, while GW--EMW interactions are examined in Sec. \ref{subsec:radiation}.

\subsection{Static electromagnetic field}\label{subsec:static}
When an EMW propagates through a static electromagnetic field, it is resonantly converted to a GW of the same frequency and wave vector; the GW is sourced by a stress--energy tensor proportional to both the radiative and static electromagnetic fields \cite{G1962}. Astrophysical GWs generated through this ``Gertsenshte\u\i n process'' are generally too weak to be of practical interest \cite{PT1972}. The Gertsenshte\u\i n effect and its inverse process---where a GW in a static electromagnetic field induces an EMW proportional to both fields---might nevertheless be relevant for detecting individual gravitons \cite{D2013} or high-frequency GWs \cite{LWF2013}.

The inverse Gertsenshte\u\i n process is as inefficient as its counterpart, and the fraction of gravitational energy converted is small ($<10^{-10}$) even under pulsar conditions \cite{Z1974}. However, the energy in the induced EMW might be comparable to that radiated conventionally by astrophysical systems where both the gravitational radiation and magnetic field are strong (but still in the far-field regime of Sec. \ref{sec:interactions}). Hence it is worthwhile to derive the inverse Gertsenshte\u\i n effect within our framework, and to revisit the feasibility of detecting it in observations.

For a plane GW propagating in a uniform magnetic field, the field component in the direction of the wave vector does not affect the induced EMW. Considering only the projection of the magnetic field onto the $xy$-plane, we have
\begin{equation}
E^{(0)}_a=0,\quad B^{(0)}_a=B^{(0)}p^{(0)}_a,
\end{equation}
with the unit polarisation vector $p^{(0)}_a=(0,\cos{\beta},\sin{\beta},0)$ for some field polarisation angle $\beta$. All linear corrections in Eqs \eref{eq:divEcorr}--\eref{eq:boxBcorr} vanish for static and uniform electromagnetic fields, and we expect separable solutions to the system. Isolating the spatial dependence in our ansatz as a scalar harmonic, we write
\begin{equation}
E^{(1)}_a=\mathcal{E}_a\exp{(ikz)},\quad B^{(1)}_a=\mathcal{B}_a\exp{(ikz)},
\end{equation}
where $\{\mathcal{E}_a,\mathcal{B}_a\}$ depends only on time.

Eqs \eref{eq:divEcorr}--\eref{eq:boxBcorr} now simplify to an ODE in time for the sole independent component of the induced field. Solving this with homogeneous initial conditions, we arrive at
\begin{equation}\label{eq:statsol}
\mathcal{E}_a=\epsilon_a^{\enspace bc}\mathcal{B}_b\delta^3_c,\quad \mathcal{B}_a=\frac{1}{2}hB^{(0)}(kt\exp{(-ikt)}-\sin{kt})p^{(1)}_a,
\end{equation}
where $h=2\sigma/k$ and $p^{(1)}_a=p_a^{\enspace b}p^{(0)}_b=(0,\cos{(2\alpha-\beta)},\sin{(2\alpha-\beta)},0)$. Eqs \eref{eq:statsol} describe a plane, monochromatic and linearly polarised EMW; its amplitude is given by
\begin{equation}\label{eq:statsolamp}
E^{(1)}=B^{(1)}=\frac{1}{2}hB^{(0)}(k^2t^2-kt\sin{(2kt)}+\sin^2{kt})^\frac{1}{2},
\end{equation}
which is proportional to time for large $t$.

The period $T_{\mathrm{GW}}=2\pi/k$ and strength $h$ of the sinusoidal background GW determine a natural timescale $T_{\mathrm{GW}}/(2\pi h)$, at which $B^{(1)}\sim B^{(0)}$ and higher-order perturbations to the electromagnetic field become significant. In reality, the linear growth in \eref{eq:statsolamp} is contingent on a steady build-up of oscillations over time, and is more of an upper bound for EMWs induced by chirp- or ringdown-type GWs with evolving frequency and/or amplitude. We incorporate such waveforms with the generalised model
\begin{equation}\label{eq:genGW}
\sigma_{ab}=\frac{1}{2}(k+\dot{k}t)h\exp{(-i(k+\frac{1}{2}\dot{k}t)t+ikz-\lambda t)}p_{ab},
\end{equation}
where the spatial dependence has been left unchanged from \eref{eq:sineGW} to maintain separability. When $\lambda=0$, Eq. \eref{eq:genGW} describes a linear chirp with constant chirp rate $\dot{f}_{\mathrm{GW}}=\dot{k}/(2\pi)$, while for $\dot{k}=0$ it gives a ringdown with damping timescale $\tau=1/\lambda$. Eqs \eref{eq:divEcorr}--\eref{eq:boxBcorr} may then be solved analytically to yield Fresnel-like integrals in the chirp case, and solutions with bounded exponential growth in the ringdown case.

The inverse Gertsenshte\u\i n effect is potentially significant in the context of compact astrophysical sources, since the induced EMW is proportional in strength to both $h$ and $B^{(0)}$. While the stable GWs from early inspirals might be conducive to resonant growth, any associated magnetic fields will have fallen off considerably where the wave zone for gravitational radiation begins; a non-optical approach must be used to study gravitational--electromagnetic interactions closer to such systems. We consider instead a typical LIGO source in an interaction region $I$ with the strongest possible GW strain $h_I$ and magnetic field strength $B_I$, i.e. at the inner edge $R_I:=c/(2\pi f_{\mathrm{GW}})$ of the local wave zone \cite{T1980}.

\begin{figure}
\centering
\includegraphics[width=0.7\columnwidth]{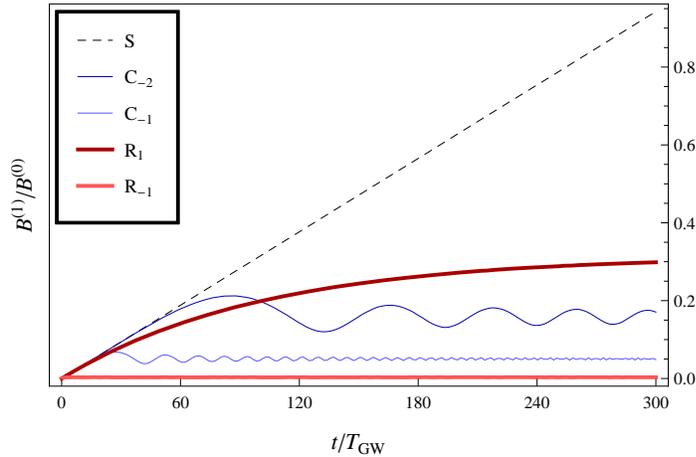}
\caption{Induced EMW amplitude relative to background magnetic field strength, for gravitational waveforms S (sinusoid), $\mathrm{C}_n$ (chirp with $\dot{f}_{\mathrm{GW}}=10^n\,\mathrm{Hz/s}$) and $\mathrm{R}_n$ (ringdown with $\tau=10^n\,\mathrm{s}$) with $f_{\mathrm{GW}}=10\,\mathrm{Hz}$ and $h_I=10^{-3}$.}
\label{fig:invgert}
\end{figure}

Fig. \ref{fig:invgert} shows the ratio $B^{(1)}/B^{(0)}$ for various gravitational waveforms, using canonical values of (initial) frequency $f_{\mathrm{GW}}=10\,\mathrm{Hz}$ and measured strain $h_\oplus=10^{-21}$ ($h_I:=h_\oplus R_\oplus/R_I=10^{-3}$) that correspond to a neutron star binary coalescence at $R_\oplus=10^2\,\mathrm{Mpc}$. With such a large interaction strain, we have $B^{(1)}\nearrow B^{(0)}$ in just 300 GW periods for the sinusoidally driven EMW, which is within the typical LIGO observation of $10^4$ waveform cycles. In general, however, the induced EMW amplitude is reduced with increasing variability in the gravitational waveform. The inverse Gertsenshte\u\i n effect is insignificant for the $\mathrm{R}_{-1}$ waveform, and hence completely negligible for actual stellar-mass ringdowns with their damping timescales of $\sim10^{-5}\,\mathrm{s}$.

Via Poynting's theorem, the spacetime-averaged power density transferred from a sinusoidal GW to its induced EMW is (to leading order in time)
\begin{equation}
-\left\langle\frac{d}{dt}u_{\mathrm{GW}}\right\rangle=\frac{1}{8\mu_0}h_I^2B_I^2\omega_{\mathrm{GW}}^2t,
\end{equation}
where units have been restored and $\omega_{\mathrm{GW}}=2\pi f_{\mathrm{GW}}$. The GW energy density is given as usual by
\begin{equation}
u_{\mathrm{GW}}=\frac{c^2}{32\pi G}h_I^2\omega_{\mathrm{GW}}^2.
\end{equation}
Hence the fraction of gravitational energy converted in the interaction region is
\begin{equation}
\Upsilon=\frac{2\pi G}{\mu_0c^2}B_I^2t^2,
\end{equation}
in accordance with the original Gertsenshte\u\i n result \cite{G1962,Z1974}. Even for a neutron star binary containing a magnetar$^2$\footnote[0]{$^2$Magnetars are highly magnetised neutron stars with typical periods of $1$ to $10\,\mathrm{s}$ and surface fields ranging from $10^9$ to $10^{11}\,\mathrm{T}$ \cite{OK2014}.} with radius $R_S=10^4\,\mathrm{m}$ and surface field strength $B_S=10^{11}\,\mathrm{T}$ ($B_I:=B_SR_S^3/R_I^3=10^3\,\mathrm{T}$), $\Upsilon$ over $10^4$ GW periods is small ($\sim10^{-9}$).

To leading order in time, the time-averaged Poynting flux of the induced EMW at the interaction distance $R_I$ is given by
\begin{equation}\label{eq:statflux}
\langle S_{\mathrm{EM}}\rangle=\frac{c}{24\mu_0}h_I^2B_I^2\omega_{\mathrm{GW}}^2t^2.
\end{equation}
Like the magnetic dipole radiation emitted by a pulsar, the Gertsenshte\u\i n radiation typically dwarfs the beamed radiation arising from synchrotron emission in the magnetosphere, but can neither propagate through the ionised interstellar medium nor be detected by existing radio telescopes due to its low frequency ($<10^3\,\mathrm{Hz}$). It is more instructive to compare \eref{eq:statflux} with the angle-averaged flux density of the maximal dipole radiation at $R_I$, which is given by \cite{P1968}
\begin{equation}
\langle S_{\mathrm{dip}}\rangle=\frac{1}{6\mu_0c^3}B_S^2\omega_{\mathrm{dip}}^4R_S^6R_I^{-2},
\end{equation}
where $\omega_{\mathrm{dip}}$ is the neutron star's angular velocity.

For a neutron star binary containing a millisecond pulsar with radius $R_S=10^4\,\mathrm{m}$ and surface field $B_S=10^6\,\mathrm{T}$, we have $\langle S_{\mathrm{EM}}\rangle\sim10^9\,\mathrm{W/m^2}$ after 300 GW periods and $\langle S_{\mathrm{dip}}\rangle\sim10^{17}\,\mathrm{W/m^2}$. If the pulsar is replaced by a similarly sized magnetar with a $1\,\mathrm{s}$ period and $10^{11}\,\mathrm{T}$ surface field, the average flux generated through the Gertsenshte\u\i n process after 300 GW periods is $\sim10^{19}\,\mathrm{W/m^2}$---a good $10^4$ times larger than that due to the magnetar's dipole radiation. Although this excess flux cannot be detected directly, it should in principle contribute significantly to the heating of any bipolar outflows or nearby interstellar clouds. The resultant secondary emission of pulsed electromagnetic radiation (with pulse frequency $f_{\mathrm{GW}}$) might then be observable by conventional telescopes across a range of bands, depending on the composition of the surrounding nebula.

\subsection{Electromagnetic radiation}\label{subsec:radiation}
Interactions between gravitational and electromagnetic radiation in the far field are most prominently characterised by a variety of interference-like (but fully nonlinear) effects on the latter. For our framework, we consider a free EMW that is plane, monochromatic and linearly polarised; since electromagnetic wavelengths are typically much shorter than gravitational and astrophysical length scales, our choice is motivated by the validity of geometrical optics as much as the suitability of plane harmonics to the tensor--vector contractions in Eqs \eref{eq:F} and \eref{eq:G}. The EMW is described by the real part of
\begin{equation}\label{eq:freeEMW}
E^{(0)}_a=E^{(0)}\exp{(i(n_bx^b+\psi))}p^{(0)}_a,\quad B^{(0)}_a=\frac{1}{n}\epsilon_a^{\enspace bc}n_b E^{(0)}_c,
\end{equation}
where the four-wave vector $n_a=n(-1,\sin\theta\cos\phi,\sin\theta\sin\phi,\cos\theta)$ has the usual polar and azimuthal angles (with respect to the $z$-direction), and $\psi$ is the initial phase relative to \eref{eq:sineGW}. The unit polarisation vector now lies in the plane orthogonal to the spatial wave vector $n_\mu$, and is defined such that $p^{(0)}_3=\sin\theta\sin\gamma$ for some wave polarisation angle $\gamma$.

For separable solutions, the tensor--vector contractions in Eqs \eref{eq:F} and \eref{eq:G} motivate the ansatz
\begin{eqnarray}\label{eq:radansatz}
\fl E^{(1)}_a=\frac{1}{2}(\mathcal{E}^{(+)}_a\exp{(im^{(+)}_\mu x^\mu)}+\mathcal{E}^{(-)}_a\exp{(im^{(-)}_\mu x^\mu)}),\nonumber\\
B^{(1)}_a=\frac{1}{2}(\mathcal{B}^{(+)}_a\exp{(im^{(+)}_\mu x^\mu)}+\mathcal{B}^{(-)}_a\exp{(im^{(-)}_\mu x^\mu)}),
\end{eqnarray}
where $m^{(\pm)}_\mu:=n_\mu\pm k_\mu$ are spatial wave vectors associated with the first-order perturbation, and we have used the phasor multiplication rule
\begin{equation}
\Re{(e^{i\Phi})}\Re{(e^{i\Psi})}=\frac{1}{2}\Re{(e^{i|\Phi+\Psi|}+e^{i|\Phi-\Psi|})}.
\end{equation}
The scalar Helmholtz harmonics $\exp{(im^{(\pm)}_\mu x^\mu)}$ decouple from \eref{eq:boxEcorr} and \eref{eq:boxBcorr}, leaving a system of ODEs in time for $\{\mathcal{E}^{(\pm)}_a,\mathcal{B}^{(\pm)}_a\}$. Although the form of \eref{eq:radansatz} is amenable to plane-wave solutions, the divergences \eref{eq:divEcorr} and \eref{eq:divBcorr} depend on the angular configuration $\{\theta,\phi,\alpha,\gamma\}$ of the waves, and are in general nonzero. As it turns out, the full system \eref{eq:divEcorr}--\eref{eq:boxBcorr} of propagation and constraint equations is inconsistent with \eref{eq:radansatz} for all but two wave configurations: parallel ($\theta=0$) and antiparallel ($\theta=\pi$), both of which yield plane-wave perturbations.

GW--EMW interactions have previously been studied in the 1+3 formalism by neglecting the electric--magnetic self-interaction terms in the propagation equations \eref{eq:boxE} and \eref{eq:boxB} (i.e. setting $G=0$ in \eref{eq:G}); this decouples the spatial dependence without explicit knowledge of the covariant Helmholtz harmonics, and for parallel waves the resultant ODE describes a resonantly driven oscillator with natural and driving frequency $m=n+k$ \cite{T2011,KT2013}. By considering the full equations, however, we find that the effect of $G$ is to cancel the terms due to $F$ when $\theta=0$, such that \eref{eq:boxEcorr} and \eref{eq:boxBcorr} become homogeneous wave equations. In other words, parallel waves do not interact at all. Such cancellation does not occur for antiparallel waves, although we find no resonant interaction either. The lack of interaction between parallel GWs and EMWs is a result that has been obtained via other approaches \cite{F1993,C1983,M1998}.

For a general interaction angle $\theta$, Eqs \eref{eq:divEcorr}--\eref{eq:boxBcorr} do not admit two-mode solutions of the form \eref{eq:radansatz}. Nevertheless, the $m^{(\pm)}$-modes are dominant when the frequency ratio $\rho:=k/n$ is small, in the sense that the propagation and constraint equations are consistent to leading order when $\rho\sec\theta\ll1$. Since $\rho<10^{-4}\ll1$ in most astrophysical scenarios (the highest-frequency GW sources have $f_\mathrm{GW}\sim10^3\,\mathrm{Hz}$ \cite{SS2009}, while $f_\mathrm{EM}\sim10^7\,\mathrm{Hz}$ is the lowest frequency that modern radio telescopes are sensitive to \cite{FEA2006,T2012}), the ansatz \eref{eq:radansatz} is justifiable and valid for all angular configurations except orthogonal waves.

Solving the time ODEs for $\{\mathcal{E}^{(\pm)}_a,\mathcal{B}^{(\pm)}_a\}$ with homogeneous initial conditions, we obtain a wave perturbation with a complicated dependence on $\{k,n,\theta,\phi,\alpha,\gamma\}$ (see Appendix). The solution \eref{eq:radsol} remains linearly polarised, however, as its components have a common phase offset $\psi$ and time dependence 
\begin{eqnarray}
\fl\mathcal{E}^{(\pm)}_a,\mathcal{B}^{(\pm)}_a\propto m^{(\pm)}\exp{(-i(n\pm k)t)}\nonumber\\
-m^{(\pm)}\cos{(m^{(\pm)}t)}+i(n\pm k)\sin{(m^{(\pm)}t)},
\end{eqnarray}
where $m^{(\pm)}=(k^2+n^2\pm2kn\cos{\theta})^{1/2}$. This represents an effective splitting of the EMW frequency into four perturbation frequencies $m^{(\pm)}$ and $n\pm k$, along with the original free frequency $n$. The amplitude of the wave perturbation vanishes in the limit for parallel waves, and is $\sim hE^{(0)}$ for antiparallel waves; its characteristic size for general $\theta$ is
\begin{equation}
E^{(1)},B^{(1)}=\mathcal{O}(hE^{(0)}/\rho),
\end{equation}
which indicates that nonlinear interference effects between GWs and EMWs become significant as $h\nearrow\rho$. When $h>\rho$, higher-order perturbations come into play and the validity of the perturbative approach might be limited.

\begin{figure}
\centering
\captionsetup[subfloat]{position=top}
\subfloat[]{\includegraphics[width=0.5\columnwidth]{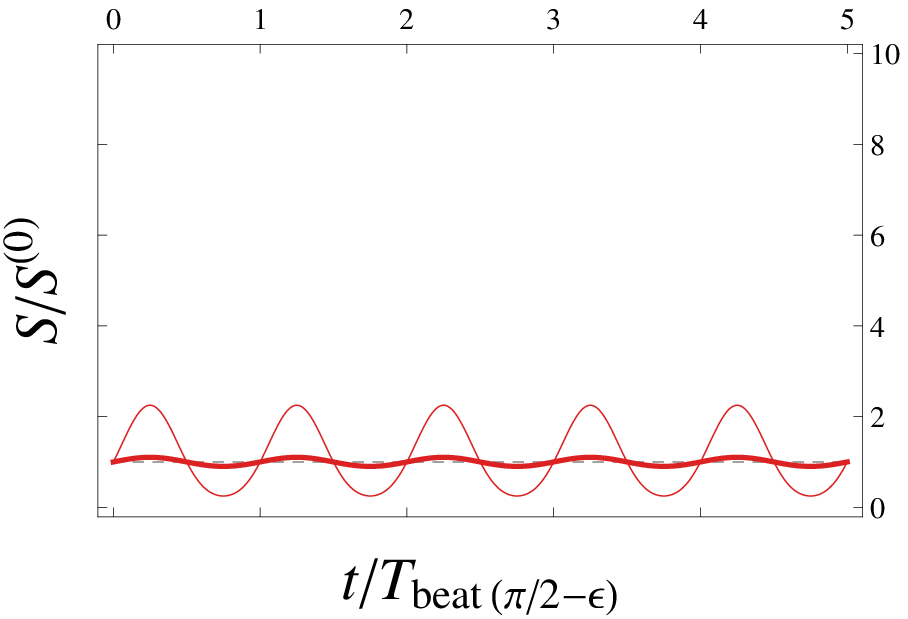}}
\subfloat[]{\includegraphics[width=0.5\columnwidth]{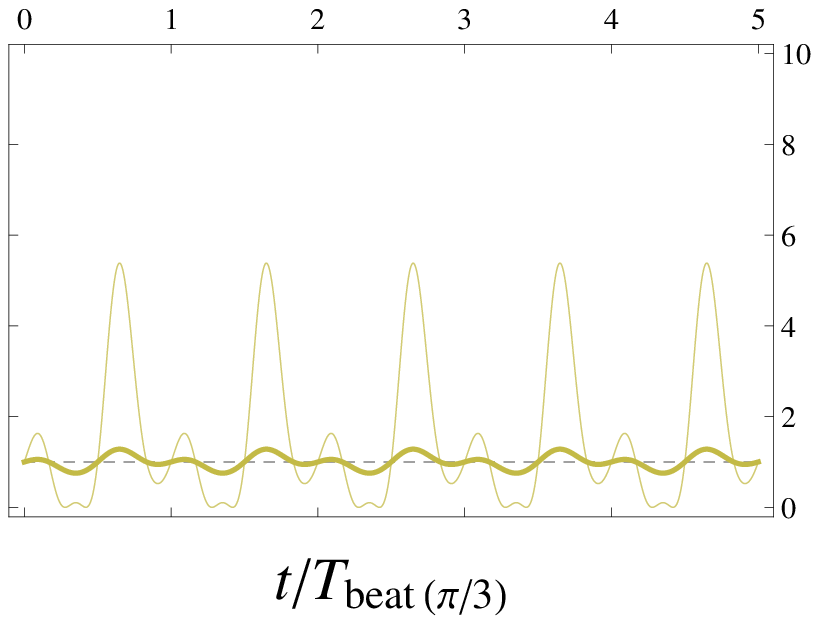}}
\\
\subfloat[]{\includegraphics[width=0.5\columnwidth]{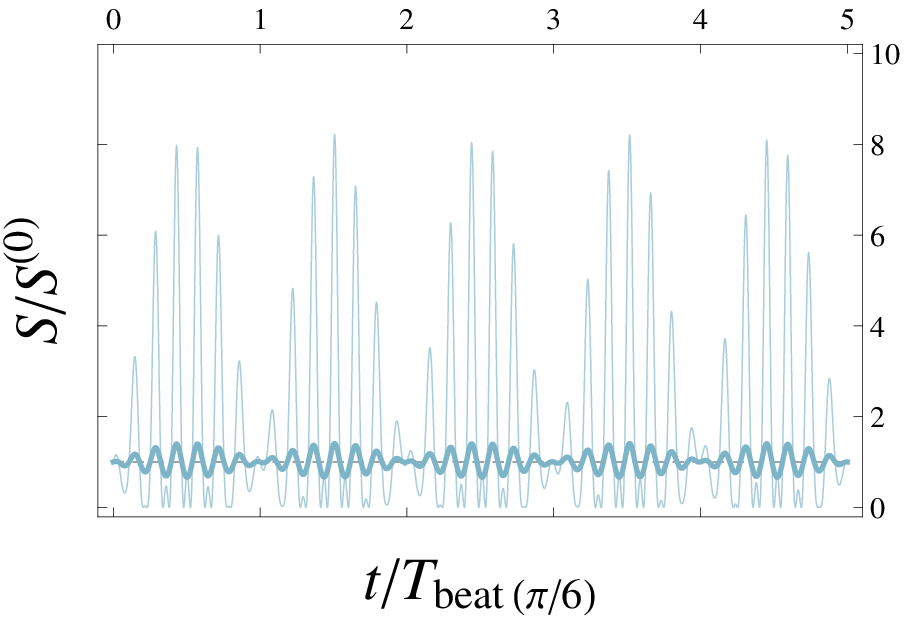}}
\subfloat[]{\includegraphics[width=0.5\columnwidth]{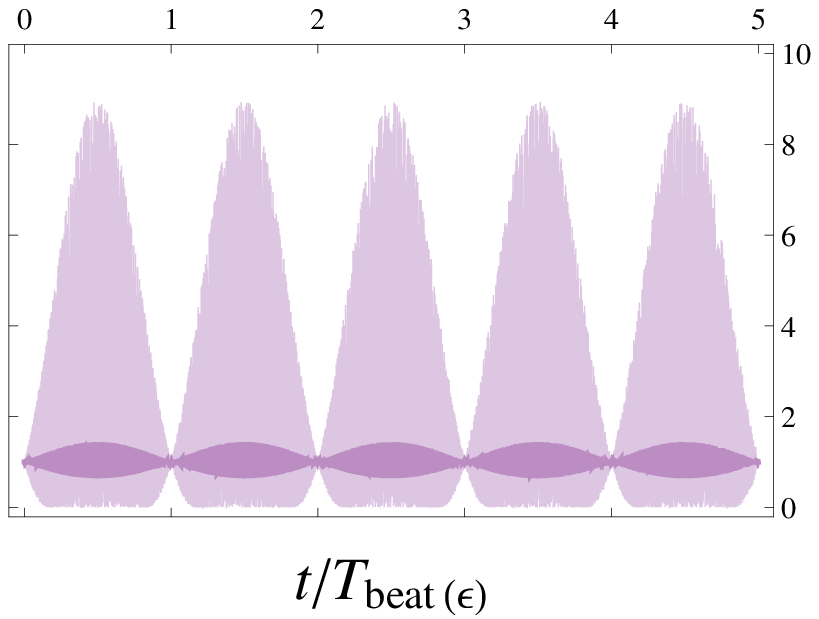}}
\\
\subfloat[]{\includegraphics[width=0.5\columnwidth]{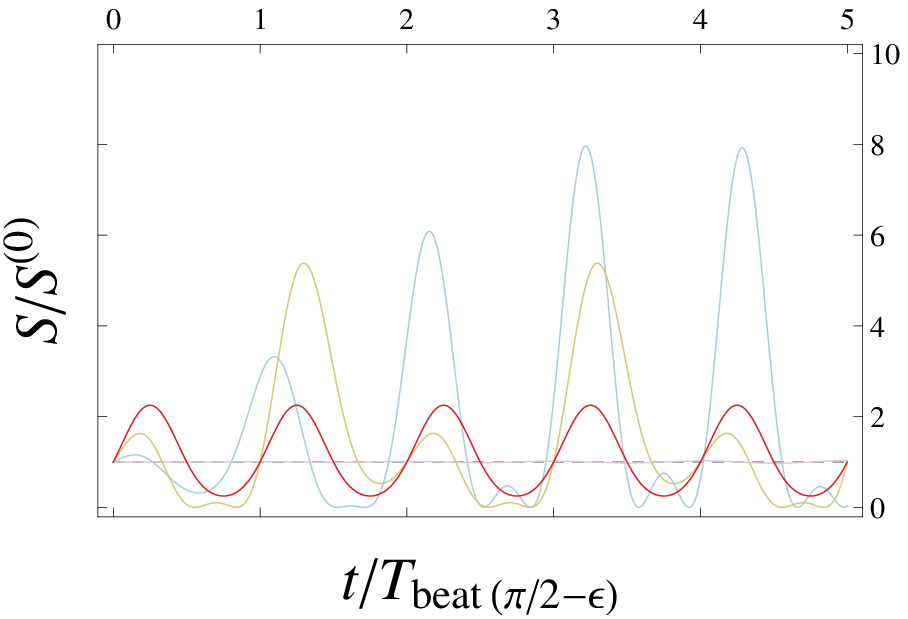}}
\subfloat[]{\includegraphics[width=0.5\columnwidth]{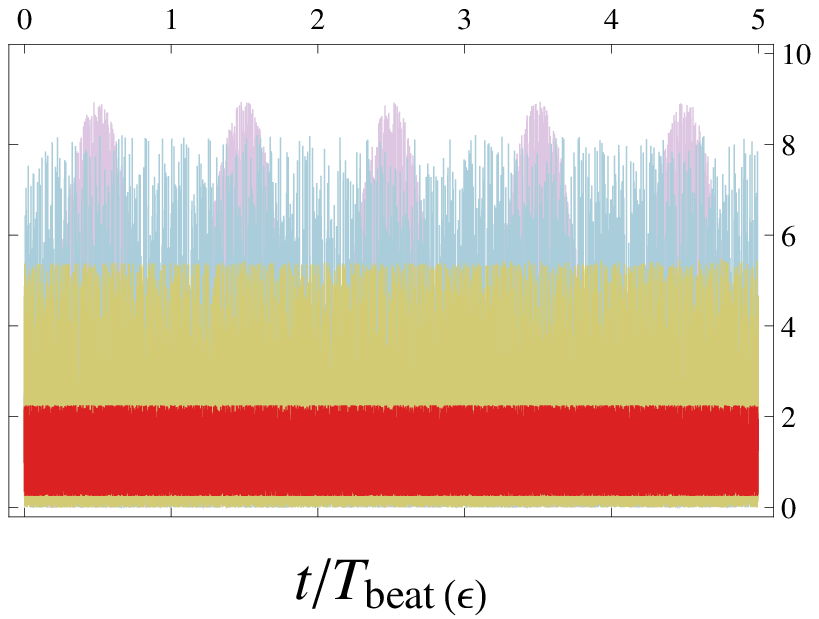}}
\caption{Perturbed Poynting flux envelope for $h=\rho$ at different interaction angles between $\theta=\pi/2$ and $\theta=0$ ((a)--(d)), along with comparisons of the four configurations over different timescales ((e) and (f)). The bolded curves in (a)--(d) are for $h=10^{-1}\rho$.}
\label{fig:flux}
\end{figure}

To illustrate the behaviour in the $h\sim\rho$ regime, we define a complex Poynting vector$^3$\footnote[0]{$^3$Our definition differs by a factor of $1/2$ from the conventional complex Poynting vector, which is used to calculate time-averaged flux for sinusoidal plane waves.}
\begin{equation}\label{eq:comPoyn}
S_a:=\epsilon_a^{\enspace bc}E_bB^*_c,
\end{equation}
which gives the envelope $S=(S^aS^*_a)^{1/2}$ of the usual Poynting vector magnitude for the full field $\{E_a,B_a\}$. Due to the presence of cross terms in Eq. \eref{eq:comPoyn}, the Poynting flux envelope is spatially periodic on the gravitational length scale $2\pi/k$. Fig. \ref{fig:flux} shows the relative flux envelope $S/S^{(0)}$ for a $+$-polarised GW ($\alpha=0$) and an EMW in the $yz$-plane ($\phi=\pi/2,\gamma=0$), where $S$ is evaluated at $x^\mu=0$ and $S^{(0)}$ is the constant flux envelope of the free EMW.

In accordance with previous results \cite{KT2013}, there is an emergence of $\theta$-dependent beats in the perturbed EMW with frequency given by the greatest common divisor of the spectrum $\{n,m^{(\pm)},n\pm k\}$. It is more useful to define an approximate beat period $T_\mathrm{beat(\theta)}=2\pi/(k-(m^{(+)}-m^{(-)})/2)$ instead, which describes much of the beat structure for most values of $\theta$. As the interaction angle decreases from $\pi/2-\epsilon$ to $\epsilon$ (where $\epsilon<10^{-3}$), the peaks for the extremal case $h=\rho$ increase from around $S/S^{(0)}=2$ to a limiting value of $S/S^{(0)}=9$. Additionally, we find significant nonlinear amplification of the beats as $h$ is raised from $10^{-1}\rho$ to $\rho$. Beating effects are essentially negligible for $h<10^{-3}\rho$.

\begin{figure}
\centering
\includegraphics[width=\columnwidth]{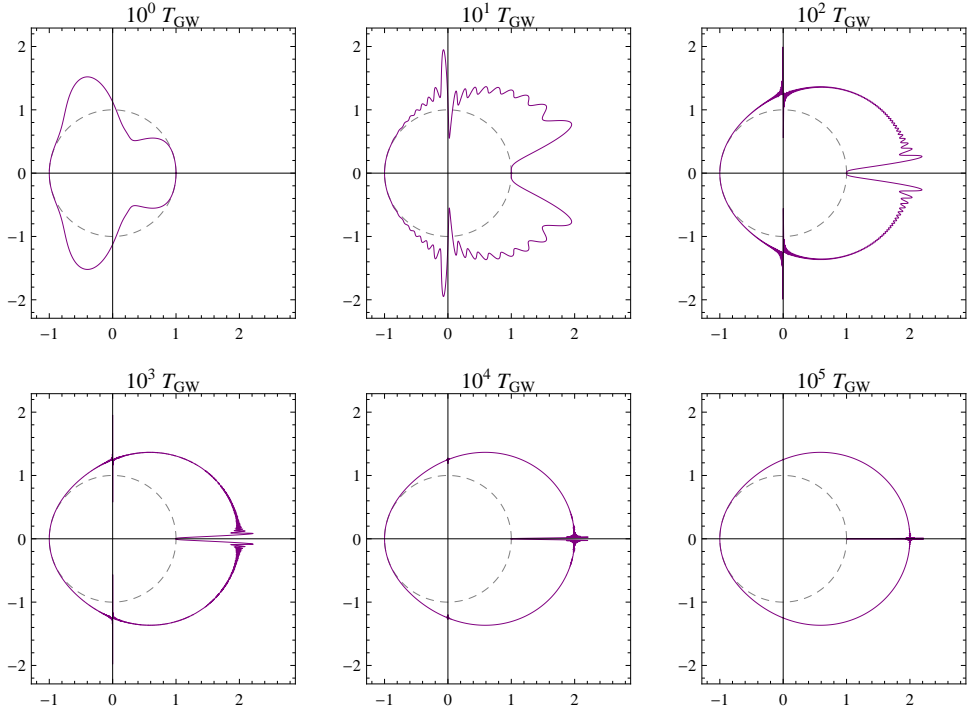}
\caption{Perturbed time-averaged Poynting flux $\langle\mathfrak{S}\rangle/\langle\mathfrak{S}^{(0)}\rangle$ over $10^0$ to $10^5$ GW periods, as a radial function of interaction angle. Each plot is for $h=\rho$, with $\theta=0$ on the positive horizontal axis such that the GW propagates to the right.}
\label{fig:fluxamp}
\end{figure}

There is an overall flux increase apparent in Fig. \ref{fig:flux}, attributable to the transfer of energy from the GW to the electromagnetic field as in Sec. \ref{subsec:static}. For a clearer picture of this flux amplification and its dependence on interaction angle, we require the time-averaged Poynting flux $\langle\mathfrak{S}\rangle:=(\langle\mathfrak{S}_a\rangle\langle\mathfrak{S}^a\rangle)^{1/2}$ over finite time intervals $T$, with the (real) Poynting vector and its time average given as usual by
\begin{equation}
\mathfrak{S}_a=\epsilon_a^{\enspace bc}\Re{(E_b)}\Re{(B_c)},
\end{equation}
\begin{equation}\label{eq:timeavg}
\langle\mathfrak{S}_a\rangle=\frac{1}{T}\int_0^T\mathfrak{S}_a\,dt.
\end{equation}

Considering the same angular configuration as before, Fig. \ref{fig:fluxamp} shows a sequence of polar plots (with respect to interaction angle) for $\langle\mathfrak{S}\rangle/\langle\mathfrak{S}^{(0)}\rangle$ averaged over increasing time intervals, where $\langle\mathfrak{S}\rangle$ is evaluated at $x^\mu=0$ and $\langle\mathfrak{S}^{(0)}\rangle$ for the free EMW is effectively constant over gravitational timescales. When $h\sim\rho$, the overall flux in the forward sector $|\theta|<\pi/2$ is approximately doubled for small $|\theta|$ after just $10^2$ GW periods. There is little to no flux amplification in the backward sector $|\theta|>\pi/2$. We note that the interaction between parallel waves vanishes as expected, with the beat frequency and the induced field itself going to zero smoothly as $|\theta|\to0$; the seemingly pathological behaviour of $\langle\mathfrak{S}\rangle/\langle\mathfrak{S}^{(0)}\rangle$ at $\theta=0$ is due to the non-smoothness of the time-averaging operation \eref{eq:timeavg} in the limit as $T\to\infty$.

The nonlinear interference depicted in Figs \ref{fig:flux} and \ref{fig:fluxamp} is potentially relevant for GW sources with electromagnetic counterparts that are long-lived (lasting at least several GW periods), and preferably high-frequency ($\rho<10^{-10}$) for effects to be significant at low GW strains. Possible counterparts for a compact binary coalescence are a pulsar component as in Sec. \ref{subsec:static} or, more promisingly, an extended electromagnetic source such as a bipolar ouflow or interstellar cloud around the binary. If an extended source emits radiation in the band $f_\mathrm{EM}\sim f_\mathrm{GW}/h_I$, its radiation profile might be characterised by intensity fluctuations and overall flux amplification at small angular distances from the binary's sky location; the fluctuations should increase in frequency to $f_\mathrm{beat(\pi/2)}=f_\mathrm{GW}$ as the interaction angle widens, then diminish rapidly at larger angular distances as $h_I$ falls below $10^{-1}\rho$.

Fig. \ref{fig:fluxamp} effectively describes the flux amplification at different interaction angles, but there is actually a tiny deflection of the perturbed time-averaged Poynting vector $\langle\mathfrak{S}_a\rangle$ in the direction of the GW. The original Poynting vector, averaged over all time, is given simply by
\begin{equation}\label{eq:avgPoynfree}
\langle\mathfrak{S}^{(0)}_a\rangle=\frac{1}{2}\Re{(S^{(0)}_a)}.
\end{equation}
Its perturbed counterpart reduces to
\begin{equation}\label{eq:avgPoyn}
\langle\mathfrak{S}_a\rangle=\langle\mathfrak{S}^{(0)}_a\rangle+\frac{1}{2}\langle\Re{(S^{(1)}_a)}\rangle+\frac{1}{2}\langle\Re{(\epsilon_a^{\enspace bc}E^{(1)}_bB^{(1)}_c)}\rangle,
\end{equation}
since both cross terms average to zero over all time. The first two terms in Eq. \eref{eq:avgPoyn} depend only on the angular configuration, while the spatial dependence in the third is negligible for $\rho\ll1$. We consider the deflection angle $\Theta_\mathrm{def}$ between \eref{eq:avgPoynfree} and \eref{eq:avgPoyn} with the same angular configuration as before; expanding the angle in powers of $h$ and $\rho$, we find
\begin{equation}\label{eq:defangle}
\Theta_\mathrm{def}=\mathcal{O}(\min{\{h^2/\rho,\rho\}}),
\end{equation}
which is valid in the forward sector but away from $\theta=0$, where $\Theta_\mathrm{def}$ goes sharply to $\pi/2$ due to the time-averaging operation.

Eq. \eref{eq:defangle} becomes $\Theta_\mathrm{def}=\mathcal{O}(h^2/\rho)$ for $h<\rho$, such that the deflection of the time-averaged Poynting vector varies with both $f_\mathrm{GW}$ and $f_\mathrm{EM}$. This is a new result, although a frequency-dependent deflection of time-averaged flux does not necessarily imply the dispersion of light by GWs. The maximal angle $\Theta_\mathrm{def}\sim h$ agrees with previous results for the ray deflection angle in different approaches \cite{F1993,RVM2003}. A hydrogen-line radio wave passing the stellar-mass binary coalescence of Sec. \ref{subsec:static} with an impact parameter corresponding to $h_I\sim\rho\sim10^{-8}$ will have its Poynting vector deflected by $\sim10^{-3}\,\mathrm{arcsec}$; this is comparable to the deflection due to conventional gravitational lensing by the same system ($\sim10^{-2}\,\mathrm{arcsec}$). Such angular deviations are too small to be observed directly, but might be amenable to microlensing techniques.

Another particularly well-documented GW--EMW interaction is the gravitational analogue of Faraday rotation experienced by an EMW in the field of a passing GW; if the projection of the EMW polarisation vector onto the GW polarisation plane is aligned with the $+$-mode, it will undergo a slight (oscillatory) rotation as long as the $\times$-mode is nonzero \cite{C1983,M1998,HG2007}. In our framework, there is indeed no rotation for a $+$-polarised GW ($\alpha=0$) and an aligned EMW ($\phi=\pi/2,\gamma=0$), since the real parts of $E^{(0)}_a$ and $E_a$ are parallel. We consider instead the rotation angle$^4$\footnote[0]{$^4$Here we use the complex fields to smooth out oscillations on the electromagnetic timescale; the angle $\Theta$ between two complex vectors $V_a$ and $W_a$ is given by $\cos{\Theta}=\Re{(V^aW^*_a)}/(VW)$.} $\Theta_\mathrm{rot}$ between $E^{(0)}_a$ and $E_a$ for a $\times$-polarised GW ($\alpha=\pi/4$) and the same EMW at $x^\mu=0$; expanding the angle in powers of $h$, we find
\begin{equation}
\Theta_\mathrm{rot}=\mathcal{O}(h),
\end{equation}
in accordance with previous results \cite{C1983,M1998}. The rotation angle also oscillates at $\sim f_\mathrm{GW}$ as expected \cite{C1983}, with beat frequency $f_\mathrm{beat(\theta)}$. Again, since $\rho<10^{-4}$ in most astrophysical scenarios and $h<\rho$, typical GW-induced rotations are $<10\,\mathrm{arcsec}$ and difficult to detect using current techniques.

\section{Conclusion}
In this paper, we have studied far-field interactions between gravitational radiation and electromagnetic fields in the 1+3 covariant approach to general relativity, with a view to characterising observable signatures on the electromagnetic radiation emitted by astrophysical GW sources. Linearised evolution and constraint equations for the electromagnetic field on a GW-perturbed spacetime have been approximated and solved perturbatively on Minkowski space, where the relevant harmonic expansions are explicitly known and analytically tractable.

We have rederived the inverse Gertsenshte\u\i n effect by applying this framework to the interaction of a plane GW with a static electromagnetic field, and considered the resonantly induced electromagnetic radiation in an astrophysical setting. Order-of-magnitude calculations have shown that the Gertsenshte\u\i n radiation is comparable to the magnetic dipole radiation for highly magnetised pulsars in compact binary systems; in the presence of a surrounding nebula, this might lead to a secondary emission of electromagnetic radiation pulsed at the GW frequency.

Several geometrical-optics effects have been found in the case of interacting GWs and EMWs. There is no resonant growth of the electromagnetic field as found in previous work, due to the additional consideration of electric--magnetic self-interaction contributions in this paper. We have also demonstrated that the nonlinear fluctuation and amplification of electromagnetic energy flux becomes significant as the GW strain approaches the GW--EMW frequency ratio from below, and might serve as a distinctive astrophysical signature of gravitational radiation emitted near or within an extended electromagnetic source.

From the various assumptions and approximations employed in this work, it is evident that the analytical advantages of the 1+3 approach are limited even for our simple model. A calculation of second-order perturbations induced by the first-order fields via Eqs \eref{eq:divEcorr}--\eref{eq:boxBcorr} might provide a clearer picture of interacting waves in the $h\sim\rho$ regime, although a rapid blow-up of the full field (signalling the breakdown of the perturbative approach) is more likely. Numerical solutions of \eref{eq:boxE} and \eref{eq:boxB} or their unlinearised versions might be worth pursuing in this case, both to verify results from the perturbative approach and to facilitate more accurate models by extending the framework into the $h>\rho$ regime.

Our results have observational implications for two types of astrophysical source: compact sources with large values of $h$ and $B^{(0)}$ (for the inverse Gertsenshte\u\i n effect to be relevant), and extended ones with a wide range of interaction angle (for more prominent nonlinear interference effects). They are not restricted to any specific example suggested here, however; neither have we considered scenarios where the gravitational and electromagnetic sources are separate. Detailed source models that incorporate far-field gravitational--electromagnetic interactions---or any Einstein--Maxwell coupling in general---will be an asset to GW detection efforts at present, and indeed the larger realm of GW astronomy in the future.

\section*{Acknowledgements}
We thank Christos Tsagas for helpful comments. AJKC's work was supported by the Cambridge Commonwealth, European and International Trust. PC's work was supported by a Marie Curie Intra-European Fellowship within the 7th European Community Framework Programme (PIEF-GA-2011-299190). JRG's work was supported by the Royal Society.

\appendix
\section*{Appendix}
\setcounter{section}{1}
For $\rho\sec\theta\ll1$ and homogeneous initial conditions, the solution to Eqs \eref{eq:divEcorr}--\eref{eq:boxBcorr} with the GW \eref{eq:sineGW}, the free EMW \eref{eq:freeEMW} and the ansatz \eref{eq:radansatz} is given by
\begin{equation}\label{eq:radsol}
\mathcal{E}^{(\pm)}_a=hE^{(0)}\xi^{(\pm)}(t)e^{i\psi}P^{(\pm)}_a,\quad\mathcal{B}^{(\pm)}_a=hE^{(0)}\xi^{(\pm)}(t)e^{i\psi}Q^{(\pm)}_a,
\end{equation}
where
\begin{eqnarray}
\fl\xi^{(\pm)}(t)=m^{(\pm)}\exp{(-i(n\pm k)t)}\nonumber\\
-m^{(\pm)}\cos{(m^{(\pm)}t)}+i(n\pm k)\sin{(m^{(\pm)}t)},
\end{eqnarray}
\begin{equation}
P^{(\pm)}_0=Q^{(\pm)}_0=0,
\end{equation}
\begin{eqnarray}
\fl P^{(\pm)}_1=\frac{i\sin^2{(\theta/2)}}{8m^{(\pm)}(k\pm n-m^{(\pm)})(k\pm n+m^{(\pm)})}\nonumber\\
(2n^2\sin{(2\alpha-\gamma-3\phi)}+6n^2\sin{(2\alpha+\gamma-3\phi)}\nonumber\\
-n^2\sin{(2\alpha-\gamma-2\theta-3\phi)}+n^2\sin{(2\alpha+\gamma-2\theta-3\phi)}\nonumber\\
+4n^2\sin{(2\alpha+\gamma-\theta-3\phi)}+4n^2\sin{(2\alpha+\gamma+\theta-3\phi)}\nonumber\\
-n^2\sin{(2\alpha-\gamma+2\theta-3\phi)}+n^2\sin{(2\alpha+\gamma+2\theta-3\phi)}\nonumber\\
-2(8k^2+10kn+3n^2)\sin{(2\alpha-\gamma-\phi)}\nonumber\\
-2n(2k+n)\sin{(2\alpha+\gamma-\phi)}-n^2\sin{(2\alpha-\gamma-2\theta-\phi)}\nonumber\\
+n^2\sin{(2\alpha+\gamma-2\theta-\phi)}-2n(k+2n)\sin{(2\alpha-\gamma-\theta-\phi)}\nonumber\\
-2kn\sin{(2\alpha+\gamma-\theta-\phi)}-2n(k+2n)\sin{(2\alpha-\gamma+\theta-\phi)}\nonumber\\
-2kn\sin{(2\alpha+\gamma+\theta-\phi)}-n^2\sin{(2\alpha-\gamma+2\theta-\phi)}\nonumber\\
+n^2\sin{(2\alpha+\gamma+2\theta-\phi)}),
\end{eqnarray}
\begin{eqnarray}
\fl P^{(\pm)}_2=\frac{i\sin^2{(\theta/2)}}{8m^{(\pm)}(k\pm n-m^{(\pm)})(k\pm n+m^{(\pm)})}\nonumber\\
(2n^2\cos{(2\alpha-\gamma-3\phi)}+6n^2\cos{(2\alpha+\gamma-3\phi)}\nonumber\\
-n^2\cos{(2\alpha-\gamma-2\theta-3\phi)}+n^2\cos{(2\alpha+\gamma-2\theta-3\phi)}\nonumber\\
+4n^2\cos{(2\alpha+\gamma-\theta-3\phi)}+4n^2\cos{(2\alpha+\gamma+\theta-3\phi)}\nonumber\\
-n^2\cos{(2\alpha-\gamma+2\theta-3\phi)}+n^2\cos{(2\alpha+\gamma+2\theta-3\phi)}\nonumber\\
+2(8k^2+10kn+3n^2)\cos{(2\alpha-\gamma-\phi)}\nonumber\\
+2n(2k+n)\cos{(2\alpha+\gamma-\phi)}+n^2\cos{(2\alpha-\gamma-2\theta-\phi)}\nonumber\\
-n^2\cos{(2\alpha+\gamma-2\theta-\phi)}+2n(k+2n)\cos{(2\alpha-\gamma-\theta-\phi)}\nonumber\\
+2kn\cos{(2\alpha+\gamma-\theta-\phi)}+2n(k+2n)\cos{(2\alpha-\gamma+\theta-\phi)}\nonumber\\
+2kn\cos{(2\alpha+\gamma+\theta-\phi)}+n^2\cos{(2\alpha-\gamma+2\theta-\phi)}\nonumber\\
-n^2\cos{(2\alpha+\gamma+2\theta-\phi)}),
\end{eqnarray}
\begin{eqnarray}
\fl P^{(\pm)}_3=\frac{in\sin{\theta}}{8m^{(\pm)}(k\pm n-m^{(\pm)})(k\pm n+m^{(\pm)})}\nonumber\\
(2(3k+n)\sin{(2\alpha-\gamma-2\phi)}+2(k-n)\sin{(2\alpha+\gamma-2\phi)}\nonumber\\
-3k\sin{(2\alpha-\gamma-\theta-2\phi)}+k\sin{(2\alpha+\gamma-\theta-2\phi)}\nonumber\\
-3k\sin{(2\alpha-\gamma+\theta-2\phi)}+k\sin{(2\alpha+\gamma+\theta-2\phi)}\nonumber\\
-n\sin{(2\alpha-\gamma+2\theta-2\phi)}+n\sin{(2\alpha+\gamma+2\theta-2\phi)}\nonumber\\
-n\sin{(2\alpha-\gamma-2\theta-2\phi)}+n\sin{(2\alpha+\gamma-2\theta-2\phi)}),
\end{eqnarray}
\begin{eqnarray}
\fl Q^{(\pm)}_1=\frac{i\sin^2{(\theta/2)}}{8m^{(\pm)}(k\pm n-m^{(\pm)})(k\pm n+m^{(\pm)})}\nonumber\\
(2n^2\cos{(2\alpha-\gamma-3\phi)}-6n^2\cos{(2\alpha+\gamma-3\phi)}\nonumber\\
-n^2\cos{(2\alpha-\gamma-2\theta-3\phi)}-n^2\cos{(2\alpha+\gamma-2\theta-3\phi)}\nonumber\\
-4n^2\cos{(2\alpha+\gamma-\theta-3\phi)}-4n^2\cos{(2\alpha+\gamma+\theta-3\phi)}\nonumber\\
-n^2\cos{(2\alpha-\gamma+2\theta-3\phi)}-n^2\cos{(2\alpha+\gamma+2\theta-3\phi)}\nonumber\\
-2(8k^2+10kn+3n^2)\cos{(2\alpha-\gamma-\phi)}\nonumber\\
+2n(2k+n)\cos{(2\alpha+\gamma-\phi)}-n^2\cos{(2\alpha-\gamma-2\theta-\phi)}\nonumber\\
-n^2\cos{(2\alpha+\gamma-2\theta-\phi)}-2n(k+2n)\cos{(2\alpha-\gamma-\theta-\phi)}\nonumber\\
+2kn\cos{(2\alpha+\gamma-\theta-\phi)}-2n(k+2n)\cos{(2\alpha-\gamma+\theta-\phi)}\nonumber\\
+2kn\cos{(2\alpha+\gamma+\theta-\phi)}-n^2\cos{(2\alpha-\gamma+2\theta-\phi)}\nonumber\\
-n^2\cos{(2\alpha+\gamma+2\theta-\phi)}),
\end{eqnarray}
\begin{eqnarray}
\fl Q^{(\pm)}_2=\frac{i\sin^2{(\theta/2)}}{8m^{(\pm)}(k\pm n-m^{(\pm)})(k\pm n+m^{(\pm)})}\nonumber\\
(-2n^2\sin{(2\alpha-\gamma-3\phi)}+6n^2\sin{(2\alpha+\gamma-3\phi)}\nonumber\\
+n^2\sin{(2\alpha-\gamma-2\theta-3\phi)}+n^2\sin{(2\alpha+\gamma-2\theta-3\phi)}\nonumber\\
+4n^2\sin{(2\alpha+\gamma-\theta-3\phi)}+4n^2\sin{(2\alpha+\gamma+\theta-3\phi)}\nonumber\\
+n^2\sin{(2\alpha-\gamma+2\theta-3\phi)}+n^2\sin{(2\alpha+\gamma+2\theta-3\phi)}\nonumber\\
-2(8k^2+10kn+3n^2)\sin{(2\alpha-\gamma-\phi)}\nonumber\\
+2n(2k+n)\sin{(2\alpha+\gamma-\phi)}-n^2\sin{(2\alpha-\gamma-2\theta-\phi)}\nonumber\\
-n^2\sin{(2\alpha+\gamma-2\theta-\phi)}-2n(k+2n)\sin{(2\alpha-\gamma-\theta-\phi)}\nonumber\\
+2kn\sin{(2\alpha+\gamma-\theta-\phi)}-2n(k+2n)\sin{(2\alpha-\gamma+\theta-\phi)}\nonumber\\
+2kn\sin{(2\alpha+\gamma+\theta-\phi)}-n^2\sin{(2\alpha-\gamma+2\theta-\phi)}\nonumber\\
-n^2\sin{(2\alpha+\gamma+2\theta-\phi)}),
\end{eqnarray}
\begin{eqnarray}
\fl Q^{(\pm)}_3=\frac{in\sin{\theta}}{8m^{(\pm)}(k\pm n-m^{(\pm)})(k\pm n+m^{(\pm)})}\nonumber\\
(2(3k+n)\cos{(2\alpha-\gamma-2\phi)}-2(k-n)\cos{(2\alpha+\gamma-2\phi)}\nonumber\\
-3k\cos{(2\alpha-\gamma-\theta-2\phi)}-k\cos{(2\alpha+\gamma-\theta-2\phi)}\nonumber\\
-3k\cos{(2\alpha-\gamma+\theta-2\phi)}-k\cos{(2\alpha+\gamma+\theta-2\phi)}\nonumber\\
-n\cos{(2\alpha-\gamma+2\theta-2\phi)}-n\cos{(2\alpha+\gamma+2\theta-2\phi)}\nonumber\\
-n\cos{(2\alpha-\gamma-2\theta-2\phi)}-n\cos{(2\alpha+\gamma-2\theta-2\phi)}),
\end{eqnarray}
with $m^{(\pm)}=(k^2+n^2\pm2kn\cos{\theta})^{1/2}$.

\section*{References}
\bibliographystyle{unsrt}
\bibliography{references}

\end{document}